\documentclass[a4paper]{jpconf}
\usepackage{graphicx}
\begin{document}
\title{Calibration of statistical methods used to constrain pulsar geometries via multiband light curve modelling}

\author{M C Bezuidenhout$^{1}$, C Venter$^{1}$, A S Seyffert$^{1}$ and A K Harding$^{2}$}

\address{$^{1}$ Centre for Space Research, North-West University, 11 Hoffman Street, Potchefstroom, 2531, South Africa

$^{2}$ Astrophysics Science Division, NASA Goddard Space Flight Center, 8800 Greenbelt Rd, Greenbelt, MD 20771, United States}

\ead{23545496@nwu.ac.za}

\begin{abstract}
Since its launch in 2008, the \textit{Fermi} Large Area Telescope (LAT) has detected over 200 $\gamma$-ray pulsars above 100 MeV. This population of pulsars is characterised by a rich diversity of light curve morphologies. Researchers have been using both the radio and $\gamma$-ray light curves to constrain the inclination and observer angles for each of these pulsars. At first, this was done using a by-eye technique and later via statistical approaches. We have also developed two novel statistical approaches that place the radio and $\gamma$-ray data on equal footing despite their disparate relative flux errors. We chose eleven pulsars from the Second \textit{Fermi} Pulsar Catalog, both old and young, and applied these new techniques as well as the by-eye technique to constrain their geometric parameters using standard pulsar models. We present first results on our comparison of the best-fit parameters yielded by each of the aforementioned techniques. This will assist us in determining the utility of our new statistical approaches, and gauge the overlap of the best-fit parameters (plus errors) from each of the different methods. Such a statistical fitting approach will provide the means for further pulsar magnetospheric model development using light curve data.
\end{abstract}

\section{Introduction}
Prior to the launch of the Large Area Telescope (LAT) aboard the \textit{Fermi} mission in June 2008, fewer than ten pulsars had been detected in the $\gamma$-ray domain [1]. Since then the LAT has discovered\footnote{https://confluence.slac.stanford.edu/display/GLAMCOG/Public+List+of+LAT-Detected+Gamma-Ray+Pulsars} more than 200 new $\gamma$-ray pulsars, enabling multi-wavelength pulsar studies for the first time [2]. 

The simultaneous use of radio and $\gamma$-ray observations has, however, typically been complicated by the comparatively low $\gamma$-ray flux through the telescope, resulting in very disparate relative uncertainties on the respective data sets. This disparity renders traditional goodness-of-fit techniques ineffective, since the radio light curves (LCs) dominate every fit, leading some researchers to prefer so-called ``by-eye methods'' when attempting to jointly fit modelled LCs to observations in both domains (e.g., [3]). Other studies (e.g., [4,5]) have sought to address this problem by artificially inflating the errors on the radio data such that the radio and $\gamma$-ray observations carry roughly equal weight in determining the optimal fit to modelled data. These fitting methods have had limited success.

More recently, a method has been proposed for rendering the respective data sets comparable for the purposes of finding best-fit model parameters in both domains simultaneously [6]. The aim of this paper is to apply this method to eleven pulsars selected from the Second \textit{Fermi} Pulsar Catalog [2] and compare the results to those found by other studies [4,5] as well as to what is found using by-eye fitting. Furthermore, a variation on the aforementioned fitting method is introduced here and applied to the same pulsars. Conclusions are drawn as to the utility of each of these fitting techniques.

\section{Geometric pulsar models}
This study uses an idealised picture of pulsars with vacuum retarded dipole magnetic field structures [7]. Here $\gamma$-ray emission is considered to originate from so-called acceleration gaps in the pulsar's magnetosphere where the density of the corotating plasma falls below the Goldreich-Julian charge density [8]. We use two geometric models that postulate the location of these acceleration gaps: the outer gap (OG, [9]) and two-pole caustic (TPC, [10]) models. For radio emission an empirical hollow-cone model [11] is assumed.

LCs are plots of a pulsar's intensity per unit solid angle. The tilt angle $\alpha$ and the observer angle $\zeta$ measured with respect to the pulsar's rotational axis are taken as free parameters, so that model LCs can be constructed using any viable ($\alpha$,$\zeta$)-pair; the fitting methods presented in this paper therefore aim to determine which combination of $\alpha$ and $\zeta$ best reproduce observed data in both radio and $\gamma$ rays concurrently.

\section{Fitting methods}

The first approach used in this paper towards finding best-fit ($\alpha,\zeta$) combinations is simple by-eye fitting of modelled LCs onto observed data. For any given pulsar, modelled LCs for each parameter pair, and in both the $\gamma$-ray and radio domains, are successively constructed and superimposed on the pulsar's observed data. We make a qualitative decision as to whether or not the experimental data are satisfactorily reproduced by the LC realisation for a given ($\alpha,\zeta$) pair, and an inclusion contour is drawn for each pulsar. This is done twice per pulsar, once using each of the OG and TPC geometric models discussed in the previous section.

The best-fit parameters are taken to be in the centre of the contour, with the errors on this estimate comprising a square encompassing all of the contour. Note that in cases where two or more disconnected closed contours appear on the map, we study the LCs for the centre of each contour and choose the best fit among these. 

This fitting method is, of course, rather subjective: applying the same method twice for the same pulsar using the same geometric model will produce two slightly differing answers. This is especially true when there are multiple ($\alpha,\zeta$) pairs plausibly replicating the observed data for a single pulsar. As such, this fitting method is not seen as an attempt at constraining pulsar geometries in and of itself, but rather as being a sanity check, or a basis for (qualitatively) judging the accuracy of more rigorous methods: if a statistical approach produces an answer far out of line with what by-eye fitting delivers, the former result is cast into doubt.

Considering the limitations of the by-eye fitting method, a more rigorous alternative is desired. This study uses a modified version of Pearson's $\chi^{2}$ test statistic, defined by the equation 
\begin{equation}
\chi^{2} = \sum_{i=1}^{n_{\rm{bins}}} \left( \frac{E_{i} - O_{i}}{U_{i}} \right)^{2}, 
\end{equation}
where $E_{i}$, $O_{i}$, and $U_{i}$ refer to the modelled (expected) intensity, observed intensity, and uncertainty on the observed intensity in the $i$th bin of $n_{\rm{bins}}$ bins of the LC respectively. 

Ideally the minimum value of $\chi^{2}$ in either model would be approximately equal to the number of degrees of freedom, $N_{\rm{dof}} = n_{\rm{bins}}-n_{\rm{parameters}}-1$, specifying a good match between modelled and observed data. However, this rarely occurs for the models used in this study, with the minimum test statistic often being far larger than $N_{\rm{dof}}$. This indicates that the models are
still somewhat rough approximations of the real phenomena. In this light, in order for constraints to be drawn, X must be normalised such that its minimum value is $N_{\rm{dof}}$. This approach works well for single-wavelength data sets, but breaks down when trying
to find minima in two wavelengths simultaneously. In our case, the relative errors on the observed $\gamma$-ray data are much larger than they are on the radio data, so that the values of $\chi^{2}$ are much smaller for the $\gamma$-ray domain than for the radio domain. 
Simply adding the test statistics of the $\gamma$ rays and radio waves together therefore creates a combined $\chi^{2}$ map which is very much radio-dominated.

Other studies [4,5] circumvented the problem of vastly differing $\chi^{2}$ values on the $\gamma$-ray and radio datasets by artificially inflating the errors on the radio observations to match those of the $\gamma$-ray observations, both for MSPs [4] and for young and middle-aged pulsars with longer periods of rotation [5]. While this method invariably performs better than simple
addition of the maps, different pulsars often require different degrees of radio error inflation, and in some cases these errors even need to be decreased to match the wide range of errors for different pulsar $\gamma$-ray LCs.

This study aims to consistently combine the respective $\chi^{2}$ maps such that each dataset is considered on equal footing. Two methods are used to achieve this, the first being a scaling approach [6] and the second being a simple multiplication approach. 
In the $\chi^{2}$ scaling method the dynamic ranges of the $\chi^{2}$ maps for each waveband are equalised before adding them, forcing each domain to carry equal weight in the determination of the best concurrent fit. For a more detailed description of this approach, see Ref.~[6]. 
At this stage no confidence contours have been implemented, as this is under investigation using a Monte-Carlo approach. The parameter constraints found using this method therefore do not yet have associated errors. The second approach followed by this study is simply multiplying the two $\chi^{2}$ maps together for each ($\alpha,\zeta$) combination without any scaling. This method does not have confidence intervals yet either.

\section{Results}
We applied the three fitting methods described in the previous section to eleven pulsars selected from the Second \textit{Fermi} Catalog [2]. Table 1 compares the parameter constraints obtained for the three fitting methods used in this study to what other studies [3,4] found for the same pulsars using the TPC model, while Table 2 makes the same comparison for the OG model. All angles are given in degrees.

\begin{center}
\textbf{Table 1. }Best ($\alpha,\zeta$) fits --- TPC model
\begin{tabular}{lllll}
 \noalign{\global\arrayrulewidth0.05cm}
\hline 
 \noalign{\global\arrayrulewidth0.025cm}
Pulsar & Independent studies $^{[4,5]}$ & By-eye fitting & $\chi^{2}$ scaling & $\chi^{2}$ multiplying\\
\hline
 J0030+0451 & (74$\pm$2, 55$^{+3}_{-1}$) & (75$\pm$8, 59$\pm$6) & (56,73) & (45,62)\\ 
 J0205+6449 & (75$\pm$2, 86$\pm$2) & (75$\pm$6, 86$\pm$4)& (77,84) & (78,84)\\
 J0437$-$4715 & (35$\pm$1, 64$\pm$1) & (30$\pm$1, 65$\pm$1) & (29,65) & (26,62)\\ 
 J1124$-$5916 & (84$\pm$2, 89$\pm$2) & (79$\pm$9, 84$\pm$6)& (87,75) & (78,84)\\
 J1231$-$1411 & (26$^{+3}_{-4}$, 69$\pm$ 1) & (47$\pm$8, 75$\pm$3) & (33,71) & (45,72)\\
 J1410$-$6132 & (19$^{+2}_{-4}$, 6$\pm$2) & (20$\pm$3, 8$\pm$3)& (10,20) & (10,20)\\
 J1420$-$6048 & (52$\pm$2, 53$\pm$2) & (61$\pm$4, 53$\pm$5)& (60,45) & (60,45)\\ 
 J1513$-$5908 & (50$\pm$2 ,54$\pm$2) & (50$\pm$8, 40$\pm$8)& (55,46) & (55,46)\\
 J1614$-$2230 & (80$^{+8}_{-20}$, 80$^{+6}_{-4}$) & (85$\pm$5, 65$\pm$5) & (36,74) & (48,83)\\
 J1833$-$1034 & (55$\pm$2, 75$\pm$2) & (54$\pm$8, 78$\pm$4)& (53,76) & (50,79)\\
 J2229+6114 & (42$\pm$2, 55$\pm$2) & (53$\pm$5, 36$\pm$6)& (46,60) & (45,59)\\
 \noalign{\global\arrayrulewidth0.05cm}  
 \hline
  \noalign{\global\arrayrulewidth0.025cm}

\end{tabular}
\end{center}
\newpage
\begin{center}
\textbf{Table 2. }Best ($\alpha,\zeta$) fits --- OG model
\begin{tabular}{lllll}
 \noalign{\global\arrayrulewidth0.05cm}  
\hline
\noalign{\global\arrayrulewidth0.025cm}
Pulsar & Independent studies $^{[4,5]}$ & By-eye fitting & $\chi^{2}$ scaling & $\chi^{2}$ multiplying\\
 \hline
 J0030+0451 & (88$^{+1}_{-2}$, 68$\pm$1) & (85$\pm$2, 67$\pm$2)& (85,68) & (56,73)\\ 
 J0205+6449 & (73$\pm$2, 90$\pm$2) & (75$\pm$5, 86$\pm$4) & (80,87) & (80,87)\\
 J0437$-$4715 & (76$\pm$1, 46$\pm$1) & (62$\pm$3, 35$\pm$5)& (27,63) & (27,63)\\ 
 J1124$-$5916 & (83$\pm$2, 88$\pm$2) & (70$\pm$7, 86$\pm$4) & (71,88) & (70,88)\\
 J1231$-$1411 & (88$\pm$1, 67$\pm$1) & (82$\pm$4, 59$\pm$5)& (33,71) & (45,72)\\
  J1410$-$6132 & (87$\pm$2, 76$\pm$2) & (4$\pm$3, 7$\pm$3) & (47,58) & (56,84)\\
 J1420$-$6048 & (55$\pm$2, 57$\pm$2) & (57$\pm$7, 58$\pm$8) & (62,45) & (58,56)\\ 
 J1513$-$5908 & (60$\pm$2,59$\pm$2) & (57$\pm$4, 44$\pm$5) & (56,48) & (56,48)\\
 J1614$-$2230 & (64$^{+8}_{-20}$, 88$^{+2}_{-5}$) & (55$\pm$7, 86$\pm$4) & (37,75) & (37,75)\\ 
 J1833$-$1034 & (65$\pm$2, 87$\pm$2) & (58$\pm$9, 79$\pm$4) & (87,65) & (87,57)\\
 J2229+6114 & (75$\pm$2, 55$\pm$2) & (66$\pm$2, 41$\pm$2) & (64,51) & (63,50)\\
 \noalign{\global\arrayrulewidth0.05cm}
 \hline
 \noalign{\global\arrayrulewidth0.025cm}
\end{tabular}
\end{center}

As an example, we plot the best-fit LCs as predicted by each fitting method superimposed on the observed \textit{Fermi} data for PSR J0030+0451 in Figure 1.

\begin{figure}[h]
\centering
\includegraphics[scale=0.4]{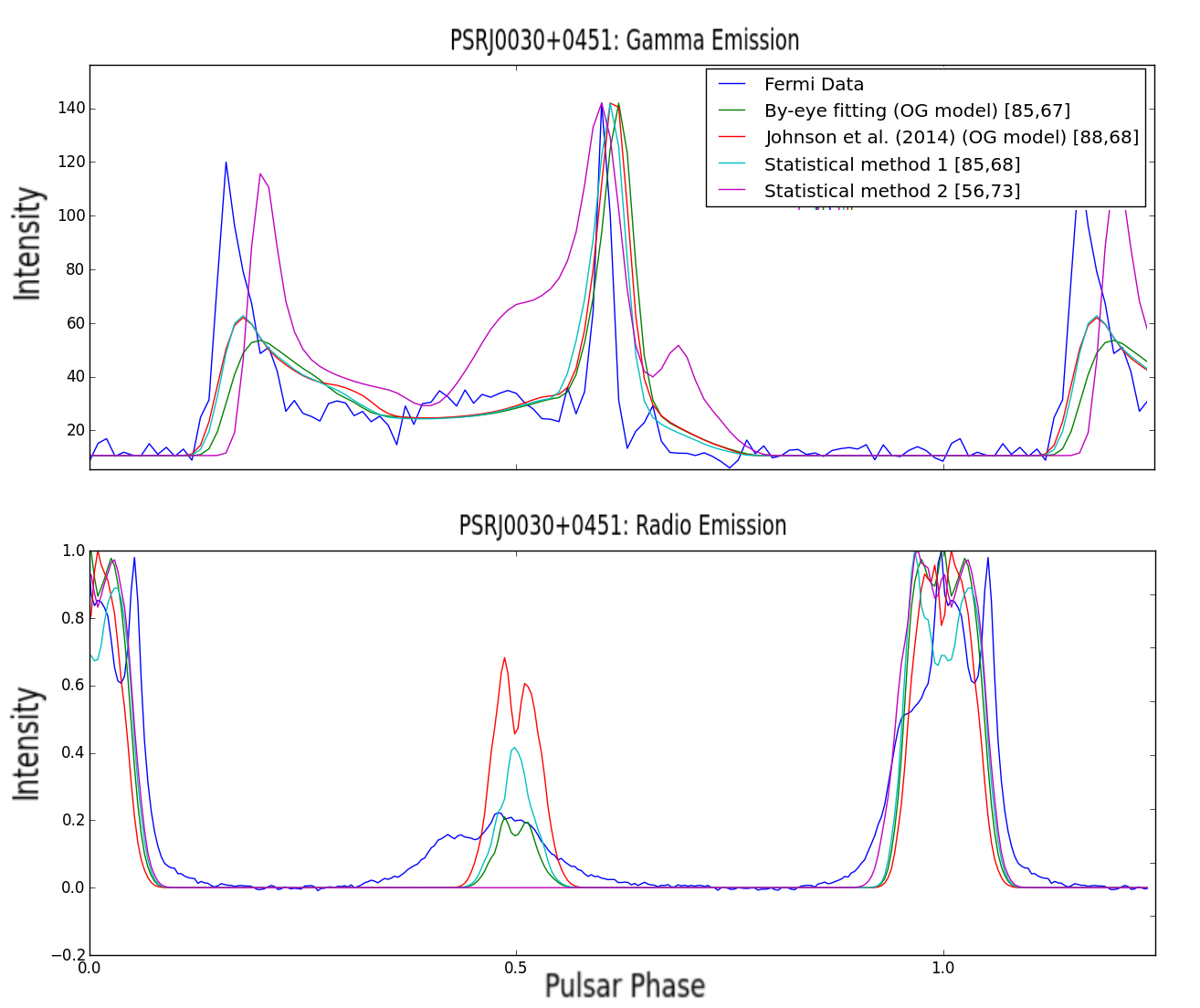} 
\caption{{\small Best-fit LCs found using each fitting method for PSR J0030+0451. The observed \textit{Fermi} LC is in dark blue, while the predicted LC found by Ref.~[4] is in red, and the LCs as predicted using the by-eye fitting, $\chi^{2}$ scaling (statistical method 1), and $\chi^{2}$ multiplying (statistical method 2) techniques are in green, cyan, and pink, respectively. The constraint pair found using each method is indicated in square brackets in the legend.}}
\end{figure}

One important effect discernible from Tables 1 and 2 is that different fitting methods often produce inverted constraint pairs, i.e., if $\alpha$ and $\zeta$ were switched, the methods would be in much better accordance. This is expected, since LC predictions are very similar for similar values of the so-called impact angle, $|\beta|=|\zeta-\alpha|$, given the model assumption of symmetric emission from both magnetic poles.

\begin{figure}[t]
\centering
\includegraphics[scale=0.75]{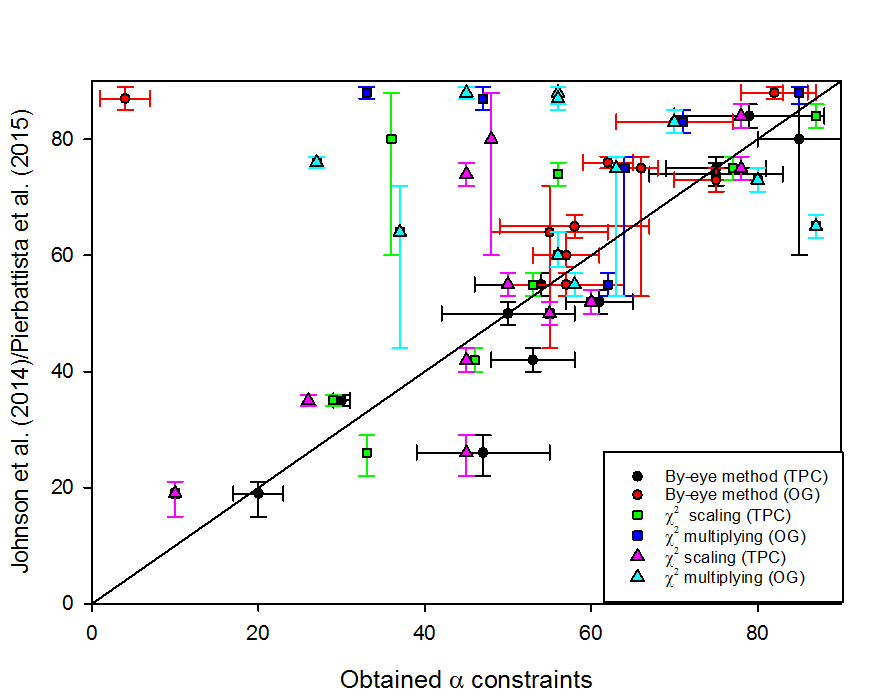} 
\caption{{\small The constraints on $\alpha$ found independently via statistical techniques vs.\ those found in this study. For reference, a straight 45$^{\circ}$ line is shown corresponding to a perfect match of the constraint we found on $\alpha$ with that found independently.}}
\end{figure}

In Figure 2 we plot the $\alpha$ constraints found in this work using by-eye fitting, $\chi^{2}$ scaling, and $\chi^{2}$ multiplying against those found by independent studies.
\section{Conclusions}

We relay the results of applying the bye-eye fitting method to the eleven pulsars in question in the third column of Tables 1 and 2. Considering Table 1 first, we find that the constraints found in this study using the TPC model overlap with those found in the independent studies in five cases. In no case would inverting the $(\alpha,\zeta)$-pair have led to concurrence between the two methods. Looking at Table 2, using the OG model, four pulsars have an overlap of intervals. Again, inverting pairs would not have produced another match.
However, the by-eye confidence intervals are fairly arbitrary, and a more lenient fitter might find significantly more overlap between these two fitting methods. The best-fit LCs found using this method also indicate that, for the most part, the by-eye fitting method is able to replicate observed LCs well, at least superficially. Taking into account the difficulties with subjectivity inherent to the by-eye fitting method, in the light of these comparisons this method seems to be a suitable basis for qualitatively judging the applicability of other, more rigorous fitting methods. 

This study's second approach to LC fitting was the $\chi^{2}$ scaling method, developed by Ref.~[6], the results of which can be found in the fourth column of Tables 1 and 2. In the TPC model, only two constraints found by this method fall inside the confidence intervals of those found by the independent studies, although PSR J0030+0451 also qualifies if its $\alpha$ and $\zeta$ are switched. There are four overlaps between the by-eye fitting and $\chi^{2}$ scaling constraints, and a further three if pair inversion is taken into account. Considering the OG model constraints, there is just one overlap (or two with pair inversion) between the $\chi^{2}$ scaling constraints and those of the independent studies; there are three overlaps with the by-eye fitting method. These comparisons are complicated by the fact that confidence intervals have not yet been implemented for the statistical methods. Even conservative errors would lead to quite a few more overlaps.

It seems clear that the $\chi^{2}$ scaling fitting method is somewhat hit-and-miss. Some observations appear to be very well fit by this method, such as PSR J1420$-$6048 using the OG model, while others are quite poorly fit, such as PSR J2229+6114 in the TPC model. 

For a blind statistical technique, the $\chi^{2}$ scaling method seems to do an adequate fitting job, although it is still some way off being a rigorous alternative to by-eye fitting. It is not clear that the constraints found using this method match those found using the by-eye method significantly more frequently than those found using the artificial error inflation technique.

The results of the third fitting approach, novel to this paper, are presented in the fifth column in each of the tables of the previous section. Regarding the constraints found using the TPC model, as in Table 1, there are no pulsars for which the constraints found using the $\chi^{2}$ multiplication technique is included in the confidence intervals of those found by the independent studies. There are five pulsars for which this method's constraints fall inside the errors of those obtained using by-eye fitting, and another if pair inversion is taken into account. Table 2 shows that in the OG model case there is one pulsar for which the $\chi^{2}$ multiplying method constraints overlap with the independent studies' constraints (two with pair inversion), and one overlap with the by-eye fitting method's constraints.

Again, the constraints found using this method do not agree with the by-eye fitting method significantly better than those found using error inflation do. This fact is reflected in the best-fit LCs plotted using this method: the degree to which observed LCs are reproduced by this method is quite variable, sometimes qualitatively better and at other times worse than what is produced by error inflation. In many cases this fitting method produces best fits identical or at least close to that of the $\chi^{2}$ scaling method. In this regard it is unclear which of these two statistical approaches produce better fits.

The most obvious avenue for the improvement best-fit LCs would be to employ more intricate geometrical models, although developing and implementing such models would be a long and arduous process. In the meantime, future study might focus on finding confidence contours in the $\chi^{2}$ fitting techniques, or on developing new methods of combining the $\gamma$-ray and radio datasets altogether, perhaps with a different test statistic.

\end{document}